\documentclass[conference]{templates/IEEEtran}
\usepackage{tabularx}
\usepackage{xspace}
\usepackage{listings}
\usepackage[underline=false]{pgf-umlsd}
\usepackage{msc}
\usepackage[htt]{hyphenat}
\usepackage{hyperref}
\hypersetup{breaklinks=true,hidelinks=true}
\usepackage{booktabs}
\usepackage{multirow}
\usepackage{color, colortbl}
\usepackage{longtable}

\def \showcomments {Show comments}
\ifdefined\showcomments
\usepackage{todonotes}
\usepackage{letltxmacro}
\LetLtxMacro{\todonote}{\todo}
\renewcommand{\todo}[2][]
{\todonote[inline, caption={#2}, size=\footnotesize, #1]
{\renewcommand{\baselinestretch}{0.5}\selectfont#2\par}}
\else
\usepackage[disable]{todonotes}
\fi

\newcommand{\taggedpara}[1]{\noindent\textbf{#1.}}

\newcommand{\sys}{\texttt{Scrappy}\xspace}

\IEEEoverridecommandlockouts
\makeatletter\def\@IEEEpubidpullup{6.5\baselineskip}\makeatother
\IEEEpubid{\parbox{\columnwidth}{
    Network and Distributed System Security (NDSS) Symposium 2024\\
    26 February - 1 March 2024, San Diego, CA, USA\\
    ISBN 1-891562-93-2\\
    https://dx.doi.org/10.14722/ndss.2024.24445\\
    www.ndss-symposium.org
}
\hspace{\columnsep}\makebox[\columnwidth]{}}

\begin{document}

\title{Scrappy: SeCure Rate Assuring Protocol with PrivacY}

\author{
\IEEEauthorblockN{Kosei Akama}
\IEEEauthorblockA{Keio University\\
akama@keio.jp}
\and
\IEEEauthorblockN{Yoshimichi Nakatsuka}
\IEEEauthorblockA{ETH Zurich\\
yoshimichi.nakatsuka@inf.ethz.ch}
\and
\IEEEauthorblockN{Masaaki Sato}
\IEEEauthorblockA{Tokai University and Keio University\\
masaaki@tsc.u-tokai.ac.jp}
\and
\IEEEauthorblockN{Keisuke Uehara}
\IEEEauthorblockA{Keio University\\
kei@sfc.keio.ac.jp}
}

\maketitle

\begin{abstract}

Preventing abusive activities caused by adversaries accessing online services at a rate exceeding that expected by websites has become an ever-increasing problem.
    CAPTCHAs and SMS authentication are widely used to provide a solution by implementing \emph{rate limiting}, although they are becoming less effective, and some are considered privacy-invasive.
    In light of this, many studies have proposed better rate-limiting systems that protect the privacy of legitimate users while blocking malicious actors.
    However, they suffer from one or more shortcomings: (1) assume trust in the underlying hardware and (2) are vulnerable to side-channel attacks.

Motivated by the aforementioned issues, this paper proposes \sys: {\bf\underline S}e{\bf\underline C}ure {\bf\underline R}ate {\bf\underline A}ssuring {\bf\underline P}rotocol with {\bf\underline P}rivac{\bf\underline Y}.
    \sys allows clients to generate unforgeable yet unlinkable \emph{rate-assuring proofs}, which provides the server with cryptographic guarantees that the client is not misbehaving.
    We design \sys using a combination of DAA and hardware security devices.
    \sys is implemented over three types of devices, including one that can immediately be deployed in the real world.
    Our baseline evaluation shows that the end-to-end latency of \sys is minimal, taking only 0.32 seconds, and uses only 679 bytes of bandwidth when transferring necessary data.
    We also conduct an extensive security evaluation, showing that the rate-limiting capability of \sys is unaffected even if the hardware security device is compromised.

\end{abstract}

\section{Introduction}
\label{sec:intro}

The invention of the World Wide Web has facilitated the hosting of many services online.
    While many services typically anticipate users to access their resources at a moderate rate, malicious users often attempt to exceed these limits.
    Consequently, online services employ techniques to slow down such users, commonly referred to as \emph{rate-limiting}.
    Several scenarios highlight the usefulness of rate-limiting users:
        \begin{itemize}
            \item \textbf{Online polls and product ratings:}
            to prevent manipulation of poll outcomes and ratings.
            \item \textbf{Services using third-party APIs (e.g., AI services, SMS text messages):}
            to prevent overuse and potential high billing. 
            \item \textbf{Services with free trials:}
            to increase the conversion rate for sustainability.
            \item \textbf{Preventing dictionary attacks:}
            to limit the number of login attempts.
            \item \textbf{Online crawlers:}
            to limit the collection of data.
        \end{itemize}

Two major techniques commonly adopted to slow down adversaries or limit their actions are SMS authentication and CAPTCHAs.
    SMS authentication authenticates users via phone numbers over the web while CAPTCHA (Completely Automated Public Turing test to tell Computers and Humans Apart)~\cite{ahn2003captcha} distinguishes between humans and a machine (i.e., bot) through puzzles that are easy for humans to solve but difficult for bots.
    Although both methods were not originally designed for rate-limit users, many services employ them for this purpose, as observed by \cite{maganis2012opaak} and \cite{nakatsuka2021cacti}.
    However, SMS authentication has privacy issues as all user actions are tied to a user's phone number, and studies have shown that CAPTCHAs degrade user experience~\cite{bursztein2010how,searles2023empirical}.

In response to these challenges, there have been proposals for rate-limiting alternatives to SMS authentication and CAPTCHAs with improved security, privacy, and efficiency.
    Opaak (OPen Anonymous Authentication frameworK)~\cite{maganis2012opaak} allows users to prove possession of a unique phone number without revealing it to the server.
    CACTI (Captcha Avoidance via Client-side TEE Integration)~\cite{nakatsuka2021cacti} is a system that allows users to generate CAPTCHA-avoiding rate-proofs proving to servers that they are not engaging in abusive actions without compromising their privacy, utilizing client-side Trusted Execution Environments (TEEs).
    CAP (Cryptographic Attestation of Personhood)~\cite{CAP:online} and PrivacyPass~\cite{PrivacyPassIETF:online, davidson2018privacypass} provide CAPTCHA alternatives by using FIDO authenticators~\cite{FIDOOverView:online} in CAP and anonymous cryptographic tokens in PrivacyPass.
However, these systems suffer from one or more of the following issues: 
    (1) Security of the system relies on the security of the underlying hardware and/or
    (2) Vulnerability to side-channel attacks.

Motivated by the aforementioned challenges, this paper presents \sys, a novel system overcoming the limitations of prior work while providing rate-limiting capabilities.
    When contacted by users, the server defines a time window $t$ and requests a \emph{rate-assuring proof} over $t$.
    In essence, rate-assuring proofs show within which time window the user sent a request to a certain server.
    These proofs are \emph{unforgeable} and, in \sys, are produced using widely available hardware security devices, acting as a source of \emph{uniqueness}.
    Malicious users attempting to produce fake proofs are easily detectable.

As with prior work, ensuring \emph{user privacy} is one of the important goals of \sys.
    This is achieved through digital signatures that remain the same for the server-defined $t$ but become \emph{unlinkable} once the rate-assuring proof uses a different $t$ or targets different servers.
    This prevents malicious servers from tracking users while blocking malicious users from accessing the server's service too frequently.

Another highlight of \sys is that its rate-limiting feature \emph{does not} rely on the security of the underlying hardware security device.
    Even if either the essential secret key proving the device's uniqueness or the secret key used to generate rate-assuring proofs is leaked, servers can still block malicious users.

\sys is \emph{agnostic} concerning the underlying hardware security device.
    We show this by implementing \sys using three different devices: TPM, hardware security token, and smartphone.
    The TPM implementation can be deployed straight out of the box, while the other implementations require a small modification to software and specifications.
    We also list devices other than the above three that can be used in \sys in Table~\ref{table:compare-unique-resource}.

To contribute to reproducible research, we have open-sourced our implementation of \sys using TPMs at \cite{ScrappyCode:online}.
    We hope this will allow future research to take advantage of our system and help reduce the friction between users and services due to rate-limiting.

\taggedpara{Contribution}
The anticipated contributions of this work are:

\begin{enumerate}
    \item The introduction of \emph{rate-assuring proofs}, a security primitive allowing servers to prevent abusive activities from users while respecting their privacy.
    \item \sys, a novel rate-limiting protocol that uses rate-assuring proofs and the Direct Anonymous Attestation protocol as building blocks and generic hardware security devices as a source of uniqueness.
    \item Three proof-of-concept implementations of \sys, with one capable of being deployed out of the box.%
    \item Contribution towards reproducible research by open-sourcing implementation of \sys over a TPM.
    \item A comprehensive analysis of the security of the protocol, notably showing some guarantee of privacy and rate-limiting even if keys are leaked.
    \item An extensive evaluation of latency, bandwidth, and storage of \sys, showing that our system is practical for real-world usage.
\end{enumerate}

\section{Related work}\label{sec:related-work}

\definecolor{Gray}{gray}{0.9}
\begin{table*}[ht]
  \caption{Comparison of the proposed method (highlighted in bold) and related work}
  \label{table:compare-related-work}
  \centering
  \begin{tabularx}{\textwidth}{p{12em} X X X X X X X}
    \toprule
    & CAPTCHA~\cite{ahn2003captcha} & SMS Auth & CAP~\cite{CAP:online} & CACTI~\cite{nakatsuka2021cacti} & Privacy Pass~\cite{davidson2018privacypass} & Opaak~\cite{maganis2012opaak} & \textbf{This paper} \\
    \midrule
    \rowcolor{Gray}
    Private key storage & - & - & SE & TEE & Undefined & *1 & \textbf{TPM} \\
    Resource for uniqueness & - & Phone number & Attestation Secret key*2 & Provisioning Key & Undefined & Phone number & \textbf{Endorsement Key} \\ 
    \rowcolor{Gray}
    Resistance to timing correlation attacks & - & - & Strong & Strong & Weak & Strong & \textbf{Strong} \\
    Rate-limiting depends on device security & No & No & Yes & Yes & No & No & \textbf{No} \\ 
    \bottomrule
  \end{tabularx}
  
  *1: Files encrypted with a master password,
  *2: Secret key installed by manufacturer

\end{table*}

This section summarizes the most relevant rate-limiting schemes.
    A side-by-side comparison of these schemes alongside the proposed method is shown in Table~\ref{table:compare-related-work}.

\subsection{Commonly used systems}

SMS authentication~\cite{SMSVerif66:online} is a method of identification and authentication using phone numbers.
    A server first sends a random, variable-length number to a preregistered phone number via SMS.
    By entering the random number into the server's website, users prove that they own the phone number.
    In addition to user authentication, servers use SMS authentication to block users who excessively access their resources.
    SMS authentication raises privacy concerns because all user actions are tied to a specific user via their phone number.
    Moreover, it is trivial for an attacker to obtain multiple phone numbers~\cite{FreeText20:online}.

CAPTCHA~\cite{ahn2003captcha} is a mechanism that distinguishes a bot from a human.
    This is achieved by challenging a user with a task that is difficult for a bot to solve but easy for a human.
    If the user fails they are determined to be a bot, and their request is dropped.
    Due to the recent advancements in machine learning, studies have demonstrated successful attacks against CAPTCHAs~\cite{george2017generative}.
    In addition, CAPTCHA farms have diminished the efficacy of CAPTCHAs~\cite{Top10Captcha:online}.
    Despite these issues, \cite{nakatsuka2021cacti} observed that CAPTCHAs are currently used as rate-limiters, raising the bar for attackers by forcing them to spend monetary and computational resources.
    However, CAPTCHAs have been criticized for reducing website usability~\cite{bursztein2010how,searles2023empirical}.

\subsection{Techniques using hardware-assisted security}

CAP~\cite{CAP:online} is a privacy-preserving CAPTCHA alternative utilizing the FIDO framework~\cite{FIDOOverView:online}.
    Users are served a random challenge and asked to use their FIDO-compatible device (i.e., authenticator) to sign that challenge.
        This makes such signatures unforgeable, as the private key is protected by the authenticator's secure element (SE)~\cite{SE}.

CACTI~\cite{nakatsuka2021cacti} is a privacy-preserving rate-limiting system that utilizes client-side trusted execution environments (TEEs)~\cite{TEEconfidential2020confidential}.
    This is facilitated via rate-proofs that allow the server to understand the number of user actions conducted in a given time window.
    CACTI maintains a counter of the number of users' actions and generates the rate-proof by signing the counter, which the server uses to check whether the counter exceeds the threshold.
    Rate-proofs are difficult to forge since CACTI stores the private key used for the digital signature within the TEE.
    Moreover, CACTI uses group signature schemes, preventing servers from linking rate-proofs to users and other rate-proofs, thereby protecting user privacy.

The limitation of CAP and CACTI is that the rate-limiting functionality depends on the security of the client-owned device.
    Protecting the secret key within the SE is crucial for CAP, as it is impossible to revoke the key, which is replicated among many FIDO authenticators~\cite{FIDO11FullBasicAttest:online}.
    This is similar to CACTI, as an adversary is able to forge rate-proofs once the secret key is extracted from the TEE.

\subsection{Cryptographic techniques}

Privacy Pass~\cite{davidson2018privacypass,PrivacyPassIETF:online} allows users to obtain anonymous cryptographic tokens each time they participate in a task (e.g., solving a CAPTCHA).
    They can then ``spend'' this token each time they encounter such a task until they run out of tokens.
    Privacy Pass protects its users' privacy by implementing this token using blind signature schemes.
However, Privacy Pass is vulnerable to timing correlation attacks, which use the time difference between the generation and usage of tokens.

Opaak (OPen Anonymous Authentication frameworK)~\cite{maganis2012opaak} provides rate-limiting for mobile phone users by employing a cryptographic primitive called periodic k-times anonymous authentication (periodic k-TAA). %
    Combining this cryptographic primitive with SMS authentication, Opaak limits the number of private keys each user can obtain, thus limiting the number of times a user can generate signatures.
There are several downsides to Opaak.
    First, Opaak assumes that a user cannot possess many phone numbers.
        This assumption is unrealistic due to services such as SMS farm services~\cite{FreeText20:online}. %
    Second, the Opaak architecture encrypts the private key using a user-defined password.
        This puts a high burden on users as they must create a password that is secure enough to withstand brute-force attacks.
    Finally, the private key gets exposed to untrusted memory.
        This makes Opaak vulnerable to attacks such as side-channel attacks.%

\section{Background}\label{sec:background}

\subsection{Group signature scheme}

Group signature schemes~\cite{chaum1991group} allow signers to prove group membership without revealing their identity by associating multiple private keys with a single public key.
    Since signatures generated using any private key can be verified by the public key, group signature schemes make it impossible for verifiers to distinguish users based on their signatures.
A group signature scheme consists of the following entities:
\begin{itemize} 
    \setlength{\itemsep}{0pt}
    \item \textbf{Group Manager (GM)} managing group members (i.e., signers) by provisioning and revoking them;
    \item \textbf{Signer} generating signatures for verifiers; and
    \item \textbf{Verifier} verifying signatures produced by the signer.
\end{itemize}

\subsection{Direct Anonymous Attestation (DAA)} \label{sec:background:daa}

\begin{figure}[t] 
  \centering
  \includegraphics[width=0.8\columnwidth]{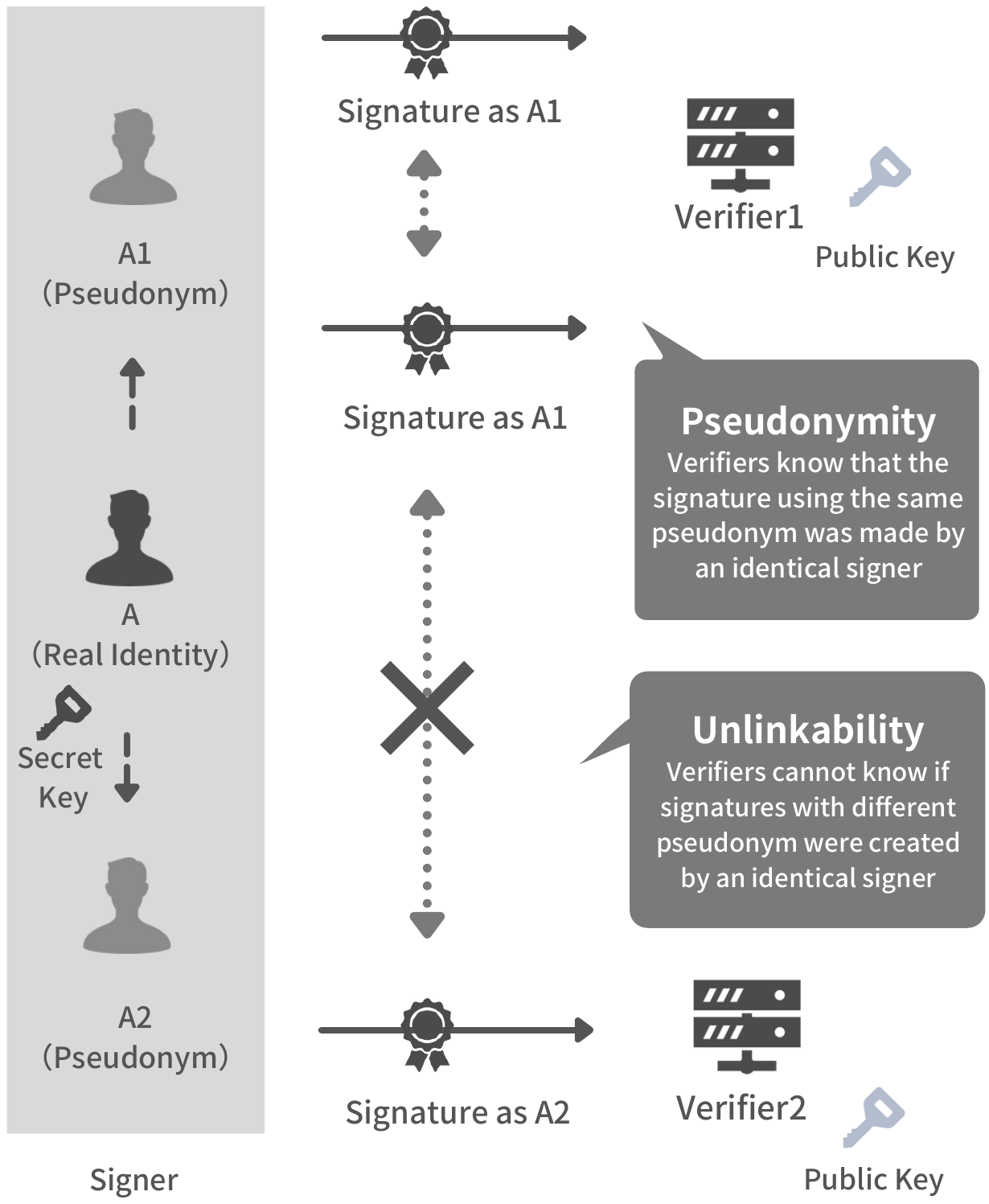}
  \caption{Pseudonymity and Unlinkability of DAA} 
  \label{fig:daa-pse} 
\end{figure} 

DAA~\cite{brickell2004direct, camenisch2013fido} (depicted in Figure~\ref{fig:daa-pse}) is a special group signature scheme that provides the following additional properties on top of those provided by group signature schemes:
(1) Neither a group member nor the GM can link a signer to a signature (\textbf{anonymity}), and
(2) Given two signatures generated for two different verifiers, it is computationally hard to determine whether they were signed by the same signer (\textbf{unlinkability}).
DAA also satisfies the \textbf{pseudonymity} property, allowing the signer to choose whether to allow multiple signatures directed to the same verifier to be linked with each other.
We briefly describe the details of the DAA protocol below using the parameters shown in Table~\ref{table:daa_param}\footnote{The details described here follow the DAA protocol~\cite{brickell2004direct} which uses TPM-specific terminology. For the ECDAA protocol, see \cite{camenisch2013fido}.}.

\begin{table}[t]
  \caption{DAA parameters~\cite{brickell2004direct}}
  \label{table:daa_param}
  \centering
  \begin{tabularx}{\columnwidth}{c X}
    \toprule
    Notation   & Description \\ \midrule
    $gpk$      & Group public key.  \\ 
    $gsk$      & Private key of the GM. \\ 
    $upk$      & Signer's public key generated by the device. \\
    $usk$      & Signer's private key generated by the device. \\
    $cred$     & Signer's credential generated by GM. \\
    $msg$      & Message to be signed. \\
    $bsn$      & Basename (chosen by Verifier or device). \\
    $\sigma_r$ & Probabilistic part of the signature, computed over $msg$ using $usk$. \\
    $\sigma_d$ & Deterministic part of the signature, computed over $bsn$ using $usk$. \\
    $\sigma$   & Signature. $\sigma = (\sigma_r, \sigma_d)$  \\ 
    $RL$       & Revocation list storing revoked private keys $usk$. \\
    $EK$       & Endorsement Key, a unique private key for each device.  \\
    $EKPub$    & EK public key, corresponding to EK. \\
    $EKCert$   & EK certificate for $EKPub$.  \\
    \bottomrule 
  \end{tabularx}
\end{table}

\taggedpara{Setup} 
GM generates $gpk$ and $gsk$.

\taggedpara{Join}
Signer obtains $usk$ and $cred$\footnote{$cred$ is treated similarly to $usk$ and therefore it is not sent to the verifier.} by following the procedure below.
\begin{enumerate}
    \item Signer generates a key pair $upk$ and $usk$.%
    \item Signer sends $upk$ and $EKCert$ to GM.
    \item GM verifies $EKCert$.
    \item GM checks that it has not seen $EKCert$ in the past.
    \item GM generates $cred$ based on $upk$ and $gsk$.
    \item GM computes ciphertext $c = Enc(cred, EKPub)$ using $EKPub$ contained in $EKCert$.
    \item GM sends $c$ to signer.
    \item Signer obtains $cred = Dec(c, EK)$.
\end{enumerate}

\taggedpara{Sign}
Signer computes $\sigma = DAA\_{Sign}(msg, bsn, usk, cred)$ over $msg$ and $bsn$, using $usk$, $cred$ and $gpk$.
$\sigma$ and $msg$ are then sent to the verifier.

\taggedpara{Verify}
Using $msg$, $bsn$, $RL$, and $gpk$, verifier checks the validity of $\sigma$ by calculating $0|1 = DAA\_Verify(\sigma, msg, bsn, RL, gpk)$.
Signer is allowed access to the requested resource if the verification returns $1$.
    
\taggedpara{Revoke}
The signer and GM revoke $usk$ as follows:
\begin{enumerate}
    \item The signer sends $usk$ to the GM.
    \item GM adds $usk$ to $RL$.
\end{enumerate}

\subsection{Periodic k-times anonymous authentication scheme (Periodic k-TAA)}
Periodic k-TAA~\cite{maganis2012opaak} is a special group signature scheme with the following features:

\begin{itemize}
    \setlength{\itemsep}{0pt}
    \item \textbf{Rate-limiting:} The signer cannot create more than one signature that is valid for a certain time period.
    \item \textbf{Anonymity:} No one, including the GM, can trace the signer from the signature.
\end{itemize}

\begin{figure}[t]
  \centering
  \includegraphics[width=\columnwidth]{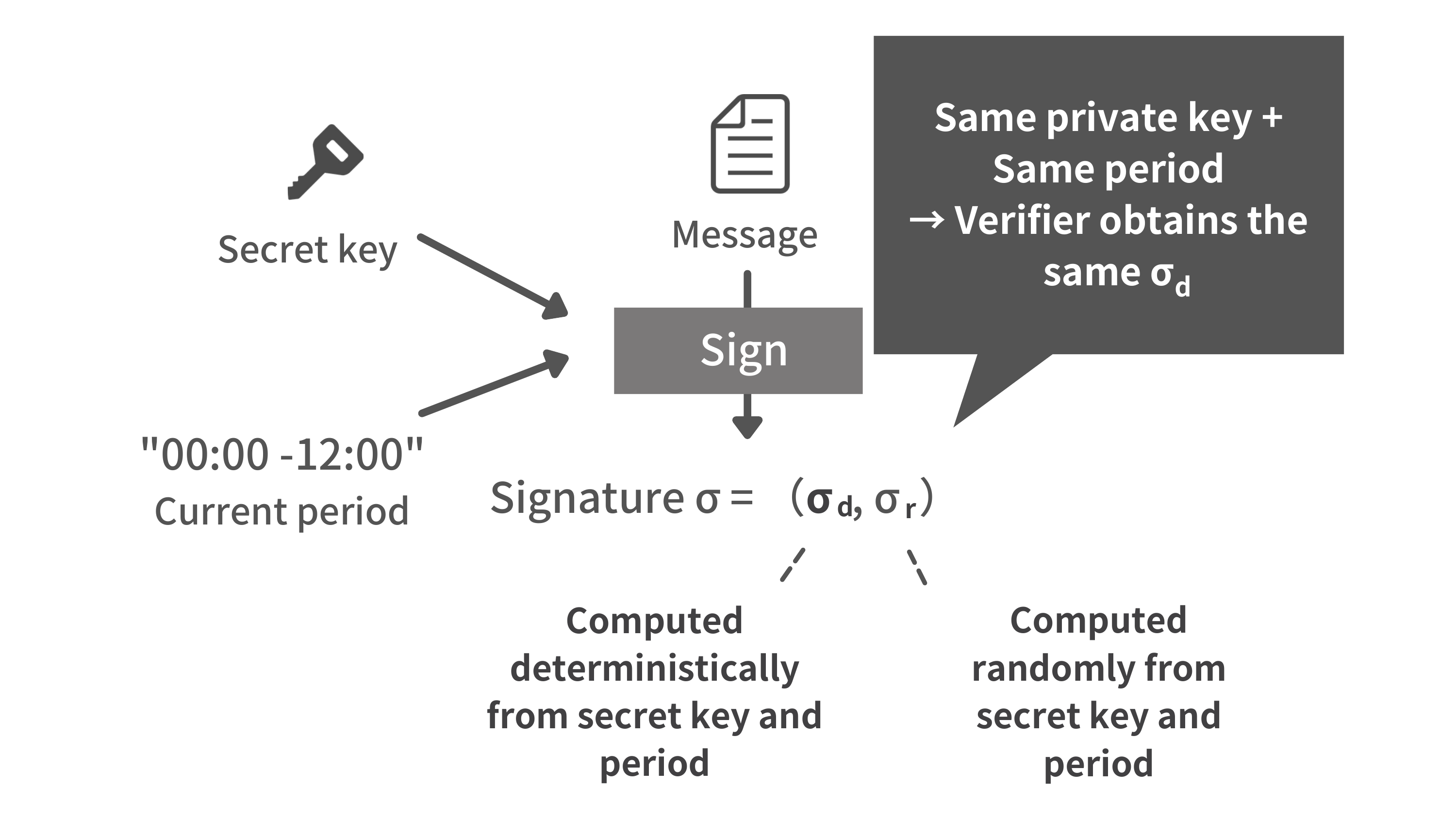}
  \caption{How periodic k-TAA works.}
  \label{fig:4_ktaa}
\end{figure}

In the periodic k-TAA scheme, depicted in Figure~\ref{fig:4_ktaa}, the verifier can determine whether the same secret key has been used to generate a signature more than once by comparing the $\sigma_d$ of the current signature with the one sent in the past.
Any $\sigma_d$ that is calculated using the same secret key during the same time window will generate the same value.

\section{System and Threat model}\label{sec:proposal}

Our proposed system consists of four types of entities: GM, signer, verifier, and hardware security device.
We assume that the number of hardware security devices is the same as the signers, and each signer owns exactly one hardware security device.
Note that a signer can always purchase multiple such devices and integrate them into their platform, but there exists a clear limit that implies higher costs to the signer.
This is discussed further in Section~\ref{sec:discussion}.
Manufacturers of the hardware security device are assumed to provide one attestation key (e.g., EK) per device.
Additionally, the GM is assumed to not provide more than one private key to the device when provisioning the device.
Finally, we assume that all cryptographic primitives are implemented correctly and that the random oracle assumption holds.

All entities other than the hardware security device are considered untrusted, i.e., they may act maliciously.
The goal of a malicious signer is to generate as many signatures as possible in a short period of time in an attempt to maximize access to resources.
The goal of a malicious verifier is to violate the signer's privacy.
The goal of a malicious GM is to collude with either the signer or verifier to help achieve their objectives.

The hardware security device acts as a source of \emph{uniqueness} and we assume it has the following properties:
    (1) Tamper-resistant memory for storing secret data,
    (2) Is capable of remote attestation,
    (3) Implements basic cryptographic primitives, including key generation, encryption/decryption, digital signatures, and
    (4) Supports a group signature scheme that provides privacy and pseudonymity.
Most importantly, \sys \emph{does not} rely on the security of the hardware security device.
    Although we assume that the device has tamper-resistant memory, we also consider powerful adversaries that can extract secret data stored in such memory.
    We show in Section~\ref{sec:security_eval} that certain properties of the proposed system are not affected even in the presence of these powerful adversaries.

Finally, we consider DoS attacks out-of-scope and do not consider malware on the signer's platform.

Our threat model yields the following security requirements for the anticipated system:
\begin{itemize}
    \setlength\itemsep{0em}
    \item \textbf{Rate-limiting:} Signers cannot send requests that exceed the verifiers' threshold.
    \item \textbf{Unforgeability:} Signers cannot forge or modify signatures.
    \item \textbf{Unlinkability:} Given two signatures generated for two different verifiers or at two different points in time, it is hard to determine whether the signatures were generated by the same signer.
\end{itemize}

We also consider the following non-security goals:
\begin{itemize}
    \setlength\itemsep{0em}
    \item \textbf{Latency:} The required latency should be minimal.
    \item \textbf{Bandwidth:} The amount of data transferred between the entities should be minimal.
    \item \textbf{Storage:} Data stored in each entity should be minimal.
\end{itemize}

\section{Design \& Challenges}\label{sec:design}

\subsection{Conceptual Design} \label{sec:concept_design}
\textbf{Rate-assuring proofs.}
In the heart of our design lies the concept of \emph{rate-assuring proofs}.
    The idea of the proof is as follows:
        The signer provides the hardware security device with a time window $t$ obtained from ``some source'' and the verifier's basename $bsn$.
            The device then generates a signature over the two pieces of data.
            This proof can be interpreted as the signer assuring the verifier that it has requested access to a specific $bsn$ within a time window $t$.
        On the other hand, the verifier keeps a ``log of proofs'' from various signers and checks whether it has received a similar proof in the past.
            If so, the verifier assumes that the signer requests resources too frequently and drops the request.

Note that rate-assuring proofs generated by a signer for \emph{different verifiers} (thus resulting in different identities) should generate indistinguishable proofs, meaning that no entity can know whether the proofs came from the same signer.
Similarly, rate-assuring proofs generated by a signer within a \emph{different time window} for the same verifier should also be indistinguishable, thus preventing adversaries from linking the two proofs to the signer.

\subsection{Design Challenges} \label{sec:design_challenges}
In realizing the design shown above, we identified the following challenges.

\textbf{Limitations in existing cryptographic primitive.}
Deciding what cryptographic approach to use to produce the rate-assuring proof is a challenge.
    The best fitting cryptographic primitive is periodic k-TAA, which meets the requirements raised in Section~\ref{sec:proposal}.
    However, there are several issues with using periodic k-TAA in tandem with hardware security devices.
        First, periodic k-TAA requires support for a special RSA modulus.
            Specifically, it uses $n = (2 p  + 1) (2 q + 1)$ instead of the widely supported $n = p \cdot q$, where $p,q$ are primes.
        Second, periodic k-TAA needs to compute $r \cdot sk + c \bmod{n}$, where $r$ is a random number, $sk$ is a secret key, $c$ is a hash, and $n$ is the aforementioned special RSA modulus.
            This calculation is used to generate Zero Knowledge Proofs, although hardware security devices do not normally support it.

\textbf{Obtaining $t$ without compromising security and privacy.}
So far, we have assumed that the hardware security device obtains $t$ from ``some source''.
    One source could be a clock on the hardware security device owned by the signer.
        However, many hardware security devices do not have a clock.
        Moreover, the clock must be synchronized frequently to prevent any skew.
            This process is difficult and, if done incorrectly, may lead to privacy leakage due to differences in $t$ caused by small skews.
    The hardware security device could also obtain $t$ from an external source.
        However, since the source may be malicious (e.g., the signer trying to cheat the system), the device must first verify the authenticity of $t$.
        This is a drawback, as it requires the hardware security device to be programmable and also puts trust in the device, which does not align with our threat model.

\textbf{Managing logs without losing functionality.}
The verifier must maintain a log of past interactions with signers for the proposed system to function correctly.
This obviously puts strain on the verifier's storage over time, as it must interact with multiple signers.

\subsection{Realizing the Design}

\begin{figure*}
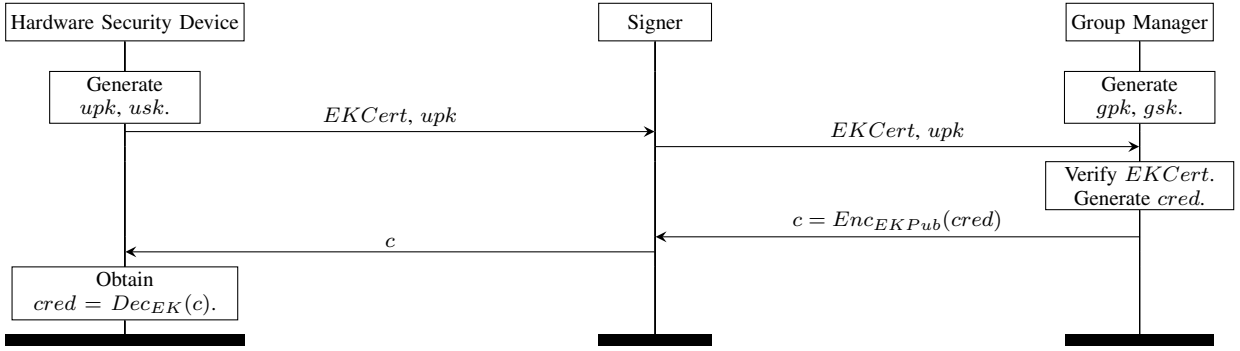

    \centering
    \begin{msc}[small values, draw frame=none, msc keyword=, head top distance=5mm, foot distance=0mm, instance distance=47mm, environment distance=10mm, level height=2mm, arrow scale=2.0, instance width=15mm, left inline overlap=15mm, right inline overlap=15mm, /tikz/font=\footnotesize, /tikz/line width=0.4pt, /msc/line width=0.4pt, label distance=0.4ex, action height=4mm]{ }
    	\declinst{t}{}{Hardware Security Device}
		\declinst{s}{}{Signer}
		\declinst{gm}{}{Group Manager}

		\action[action width=20mm]{Generate $gpk$, $gsk$.}{gm}
		\action[action width=20mm]{Generate $upk$, $usk$.}{t}
		\nextlevel[4]
		\mess{$EKCert$, $upk$}{t}{s}
		\nextlevel[1]
		\mess{$EKCert$, $upk$}{s}{gm}
		\nextlevel[1]
		\action[action width=25mm]{Verify $EKCert$.\\Generate $cred$.}{gm}
		\nextlevel[5]
    	\mess{$c = Enc_{EKPub}(cred)$}{gm}{s}
		\nextlevel[1]
    	\mess{$c$}{s}{t}
		\nextlevel[1]
		\action[action width=30mm]{Obtain $cred = Dec_{EK}(c)$.}{t}
		\nextlevel[3]
    \end{msc}
    \caption{Initialization Phase of \sys protocol. $Enc_{k}(m)$ and $Dec_{k}(c)$ denotes a cryptographic encryption/decryption operation over $m$/$c$ using key $k$, respectively.}
    \label{fig:init_phase}
\end{figure*}

\begin{figure*}
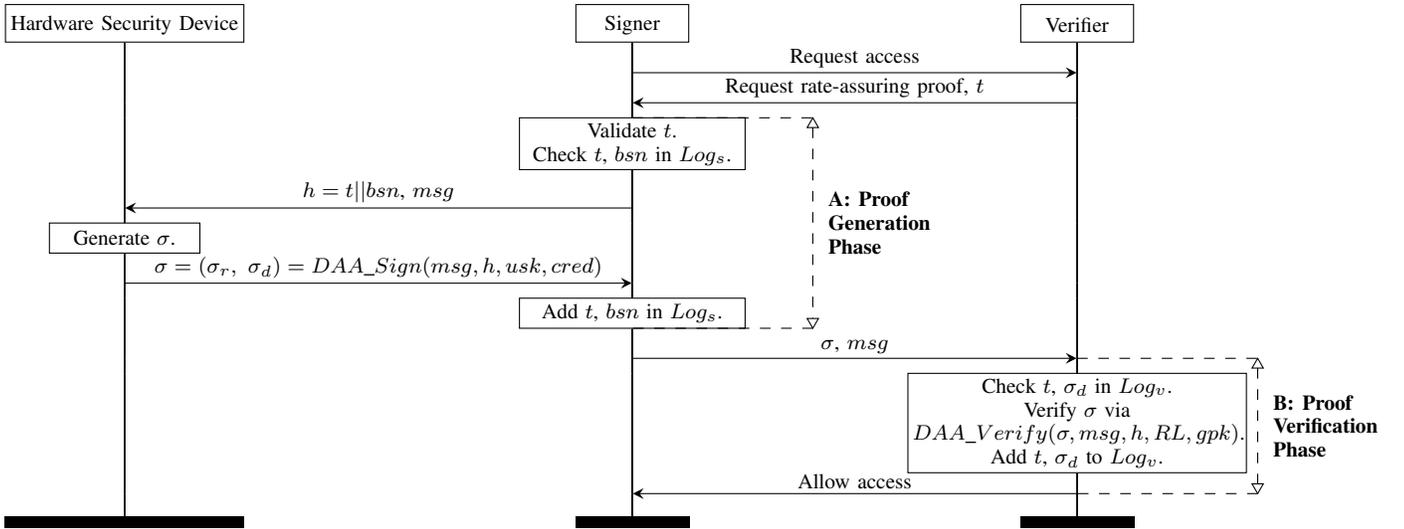

    \centering
    \begin{msc}[small values, draw frame=none, msc keyword=, head top distance=5mm, foot distance=0mm, instance distance=44mm, environment distance=10mm, level height=2mm, arrow scale=2.0, instance width=15mm, left inline overlap=15mm, right inline overlap=15mm, /tikz/font=\footnotesize, /tikz/line width=0.4pt, /msc/line width=0.4pt, label distance=0.4ex, action height=4mm]{ }
    	\declinst{t}{}{Hardware Security Device}
		\declinst{s}{}{Signer}
		\declinst{v}{}{Verifier}

		\mess{Request access}{s}{v}
		\nextlevel[2]
		\mess{Request rate-assuring proof, $t$}{v}{s}
		\nextlevel[1]
  
    	\measure[side=right, measure distance=24mm, label distance=2mm]{\parbox{20mm}{\textbf{A: Proof\\Generation Phase}}}{s}{s}[14]
		\action[action width=30mm]{Validate $t$.\\Check $t$, $bsn$ in $Log_s$.}{s}
		\nextlevel[6]
		\mess{$h = t || bsn$, $msg$}{s}{t}
		\nextlevel[1]
		\action[action width=20mm]{Generate $\sigma$.}{t}
		\nextlevel[4]
    	\mess{$\sigma = (\sigma_r,~\sigma_d) = DAA\_Sign(msg, h, usk, cred)$}{t}{s}
		\nextlevel[1]
		\action[action width=30mm]{Add $t$, $bsn$ in $Log_s$.}{s}
		\nextlevel[4]
    	\mess{$\sigma$, $msg$}{s}{v}
    	
    	\measure[side=right, measure distance=24mm, label distance=2mm]{\parbox{20mm}{\textbf{B: Proof\\Verification Phase}}}{v}{v}[9]
    	\nextlevel[1]
		\action[action width=45mm]{Check $t$, $\sigma_d$ in $Log_v$.\\Verify $\sigma$ via \\ $DAA\_Verify(\sigma, msg, h, RL, gpk)$.\\Add $t$, $\sigma_d$ to $Log_v$.}{v}
    	\nextlevel[8]
    	\mess{Allow access}{v}{s}
    \end{msc}
    \caption{Proof Generation and Verification Phases of \sys protocol. $DAA\_Sign()$ and $DAA\_Verify()$ are the DAA-specific operations described in Section~\ref{sec:background:daa}.}
    \label{fig:proof_gen_verif_phase}
\end{figure*}

This section presents a detailed design that addresses the aforementioned design challenges.
    
\textbf{Cryptographic primitive.}
Instead of the periodic k-TAA scheme, we chose to use DAA to realize \sys.
    This is because it has been the standard authentication scheme for TPMs, enabling it to perform well even on resource-constrained hardware security devices.
    Another key factor for choosing DAA is that other hardware security devices also use a similar scheme.
        For instance, FIDO has made the ECDAA specification~\cite{camenisch2013fido} publicly available, allowing future FIDO devices to use DAA during attestation.
        Additionally, Intel SGX~\cite{costan2016intel}, a widely available TEE, uses a similar primitive called EPID~\cite{brickell2007enhanced}, which could also be used instead of DAA\footnote{Although Intel has officially announced to deprecate SGX from consumer-grade CPUs~\cite{IntelSGXDeprecation:Online}, there is a possibility that the CPUs may still support EPID, allowing them to be used for Scrappy.}.

However, DAA does not provide the most important \emph{rate-limiting} property, i.e., it cannot be directly used to implement rate-assuring proofs.
    To overcome this issue, we modify the input to the $DAA\_Sign$ function to cryptographically bind $t$ to the pseudonym used to generate the signature.
        This allows verifiers to link multiple signatures generated during the same $t$ via the pseudonym.
        Note that once the $t$ changes, the signature also changes and cannot be linked to the previous one, providing unlinkability between two time periods.
    We describe how this works in detail in Section~\ref{sec:protocol}.

\textbf{Obtaining $t$.}
In the proposed system, the signer obtains $t$ from the verifier.
    A malicious verifier may provide an invalid $t$ or a maliciously crafted $t$ to track the signer (see Section~\ref{sec:security_eval} for more discussion).
    Therefore, the signer \emph{must} check whether the $t$ is valid.
    Similarly, a malicious signer may ignore the $t$ provided by the verifier and use a fake one to cheat the system.
    Therefore, the verifier \emph{must} check whether the $t$ included in the rate-assuring proof is valid.

\textbf{Managing logs.}
To reduce the impact on the verifier-side storage, we designed \sys so that verifiers can delete old entries that are no longer useful.
    In simple terms, any entries created before the current $t$ can be deleted without hindering the system's functionality.

In addition to the verifier, the signer maintains a log of past $t$ received from the verifier.
    This is to allow the signers to check whether they have produced a rate-assuring proof within the same time window and prevent them from generating rate-assuring proofs that the verifier may reject.
    \sys is designed so that the signer can also delete old records in its log to reduce storage.

\subsection{The \sys Protocol} \label{sec:protocol}

The sequence diagram of the \sys protocol is shown in Figures~\ref{fig:init_phase} and~\ref{fig:proof_gen_verif_phase}.
    We assume that the communication channels between the signer and verifier are protected via standard means, such as TLS.
    This means that the verifier holds valid certificates signed by trusted CAs that bind their domain name to a certain public key.
In addition to the parameters shown in Table~\ref{table:daa_param}, the proposed scheme uses three additional parameters:
\begin{itemize}
    \setlength\itemsep{0em}
    \item $t$: A verifier-defined time window that encompasses the current time of signing (e.g., \texttt{10:00-11:00}). The time window is inclusive at the start and cannot overlap with prior or future time windows.
    \item $Log_s$: A list of pair ($t$, $bsn$) received from the verifier in the past, i.e., $Log_s = \{(t_1, {bsn}_1), ... , (t_n, bsn_n) \}$ where $n$ is the number of items in the list.
    \item $Log_v$: A list of pair ($t$, $\sigma_d$) received from the signer in the past, i.e., $Log_v = \{(t_1, \sigma_{d1)}), ... , (t_n, \sigma_{dn} \}$ where $n$ is the number of items in the list.
\end{itemize}

\sys consists of the following four phases: (1) initialization, (2) proof generation, (3) proof verification, and (4) cleanup.
In this section, we describe each phase in detail.

\textbf{Initialization phase.}
During this phase, the GM\footnote{The GM would be required to maintain a list of certificates of trusted hardware security device vendors, which is used during verification of attestation proofs. It is worth noting that the number of vendors is expected to be small (26 for TPM~\cite{TCGTPMVendorID:online}).}, verifier, and signer initialize various variables as follows.
\begin{itemize}
    \setlength\itemsep{0em}
    \item \textbf{GM:} generates $gpk$ and $gsk$.
    \item \textbf{Verifier:} initializes its log $Log_v = \{\}$.
    \item \textbf{Signer:} generates $usk$ and $upk$, obtains $cred$ from GM, and initializes its log $Log_s = \{\}$.
\end{itemize}

\textbf{Proof generation phase.}
After the signer requests access to the verifier's resource, the verifier responds by requesting a rate-assuring proof over a $t$.
The proof is generated using $t$, $bsn$, $usk$, and $cred$, following the process shown below.

\begin{enumerate}
    \item Check whether $t$ is valid, e.g., check that it covers the current time. Abort signing if not.
    \item Obtain $bsn$ from the domain included in the verifier's certificate and check its validity. Abort signing if not.
    \item Check whether the pair $(t,~bsn)$ exists in $Log_s$. If so, abort signing.
    \item Compute value $h = bsn~||~t$.
    \item Forward $msg$ and $h$ to hardware security device to generate $\sigma$\footnote{In the actual implementation, $msg$ is represented as an empty string. Nevertheless, we leave $msg$ present here in the design, anticipating that it may be used as an extension in the future.}:
          $$\sigma = (\sigma_{r}, \sigma_{d}) = DAA\_{Sign}(msg, h, usk, cred)$$
    \item Add new pair $(t,~bsn)$ to $Log_s$.
    \item Send $\sigma$ and $msg$ to verifier.
\end{enumerate}

\textbf{Proof verification phase.}
Upon receiving $\sigma$ and $msg$ from the signer, the verifier proceeds with the verification phase by following the below process.
\begin{enumerate}
    \item Check whether $t$ is valid and reject proof if not.
    \item Check whether the pair $(t,~\sigma_d)$ exists in $Log_v$. If so, the signer has reached the rate limit, and the request must be rejected.
    \item Compute value $h = bsn~||~t$.
    \item Verify $\sigma$ via $DAA\_{Verify}(\sigma, msg, h, RL, gpk)$. If verification fails, reject the signer request.
    \item Add pair $(t,~\sigma_d)$ for the verified signature to $Log_v$.
    \item Return success and allow access to the signer.
\end{enumerate}

\textbf{Cleanup phase.}
To reduce the storage overhead of the signer and verifier, the two entities periodically ``clean up'' their logs.
    Signers delete pairs $(t,~bsn)$ for a certain verifier $bsn$ that are older\footnote{``Old'' time windows are those no longer encompassing current time.} than the newest $t$.
    On the other hand, verifiers delete all entries as soon as the current $t$ expires.

\section{Implementation}\label{sec:impl}

We implemented a Proof of Concept (PoC) of the proposed design presented in the previous section.
    The baseline implementation consists of browser extension, signer application, TPM (see Section~\ref{sec:other_devices} for implementations over additional devices), and verifier server.

All entities use SHA256 \cite{sha256} for the hash function and BN256 \cite{TCGAlgorithmRegistry:online} for the pairing curve.
    Note that BN638 \cite{TCGAlgorithmRegistry:online} is recommended over BN256 due to security reasons \cite{SecurityECDAA74:online}, and the only reason we use BN256 is that it is the only pairing curve supported by our TPM.

The verifier is set to allow one access per minute.
    Therefore, all $t$ is expressed as Unix timestamps, rounded up to the nearest minute.
    Note that this can be changed to adopt other formats of $t$. 

The implementation of \sys and the ECDAA library can be found at \cite{ScrappyCode:online}.
    Note that this release only includes the TPM implementation %
    and we plan to release the other implementations in the future.

\subsection{Browser extension}
The browser extension of the PoC is built for the Chrome browser (version 100.0.4896.60)~\cite{GoogleCh87:online}.
    Note that many browsers provide similar interfaces, allowing our extension to be easily ported to other browsers.
    The browser extension consists of 2 scripts: content script and background script.

\textbf{Content script:} responsible for obtaining the data sent from the verifier.
To do this, it looks for a special HTML tag in the verifier's HTML file.
The script parses this tag, obtains the contained parameters and the verifier's origin, and finally sends all the data to the background script.
For the PoC, we use the following implementation: \texttt{<input period="t" value="" />}, where $t$ is the server-defined time window.

The content script also sends the data received from the background script to the server.
We use the same HTML tag shown above for this purpose.
Specifically, the parameters of the HTML tag are filled in as follows: \texttt{<input period="t" value="$\sigma$" />}, where $t$ is the period provided by the verifier, and $\sigma$ is the signature generated by the hardware security device.

\textbf{Background script:} forwards the data obtained by the content script to the signer application.
    This script is necessary because the content script is designed so that it cannot directly communicate with external applications.
    The communication between the background script and the signer application uses the standard I/O and follows the Chrome Native Messaging Protocol~\cite{NativeMessageProtocol91:online}.

\subsection{Signer application}
The signer application running on the signer is responsible for (1) generating the credential by cooperating with the TPM and GM, (2) producing the rate-assuring proof with the help of the TPM using the data received from the browser extension, (3) storing the period $t$ and signature $\sigma_d$ in its log, and (4) returning the proof to the browser extension.

All signer application implementations are implemented in Golang~\cite{TheGoPro41:online} due to its high performance and memory safety.
    In addition, we used the AMCL (MIRACL/core)~\cite{amclDoc, miraclco28:online} library for cryptographic operations.
    To communicate with the TPM, the signer application uses a modified version of the \texttt{go-tpm} library~\cite{googlego0:online}, which includes support for the \texttt{TPM2\_Commit} command~\cite{TCGAlgorithmCommands:online}.

We chose to use an SQLite~\cite{AboutSQL84:online} database to implement the logging system.
    This database has a table with one column named \texttt{BASENAME}, which stores $bsn$ and $t$ as VARCHAR (56 bytes).
        To fix the size of the otherwise variable data, the signer application calculates $hash(bsn) || $``\_''$ || t$ and stores the result in the database.

Before the signer application generates the rate-assuring proof, it queries this database to check whether a proof for the same period $t$ and origin $bsn$ has been created in the past.
After the rate-assuring proof has been generated, the signer application also makes sure to update the $t$ for the $bsn$.

\subsection{TPM} \label{sec:tpm_impl}

TPM was our first choice for the hardware security device because it supports DAA by default and does not require any hardware modifications to be used for \sys.
    This enables \sys to be deployed in the real world, allowing both users and services to immediately take advantage of the various functionalities provided by \sys.
    Moreover, TPM on client-side devices is becoming more widely adopted, especially since Microsoft officially announced that the Windows 11 OS requires a TPM~\cite{windows11:online}, and we envision that more devices support TPMs (or similar cryptographic hardware) as time progresses.
    We refer the reader to Section~\ref{sec:other_devices} for implementations using devices other than TPM.

We used the ST33TPHF2ESPI TPM manufactured by ST Microelectronics~\cite{microelectronics2019st33tphf2espi}.
    The TPM clock speed is estimated to be around several tens to hundreds of MHz.
    The TPM is housed within a machine owned by the signer and generates rate-assuring proofs upon receiving requests from the signer application.
The signer application uses the following TPM functions~\cite{TCGTPMArch:online, TCGAlgorithmCommands:online} when generating the credential or the proof:
    \texttt{TPM2\_CreatePrimary}, \texttt{TPM2\_NVReadPublic}/\texttt{TPM2\_NVRead}, \texttt{TPM2\_AcativateCredential}, \texttt{TPM2\_Create}, and \texttt{TPM2\_Commit}/\texttt{TPM2\_Sign}\footnote{Although FIDO specification adopts \texttt{TPM2\_Certify}, we use \texttt{TPM2\_Sign} instead of \texttt{TPM2\_Certify}, as it supports any message type~\cite{xaptumec72:online}.}.

\subsection{Verifier server}
The verifier server conducts the following: (1) Sends the data necessary for the rate-assuring proof to the signer by embedding it into the HTML tag, and (2) verifies the proof sent by the signer.
Similar to the signer application, the server was written using Golang and uses the \texttt{gin} web framework~\cite{GinWebFr33:online}.
SQLite stores the logs and AMCL (MIRACL/core) for cryptographic operations.
    The database has a table with one column named ``K'' as VARCHAR (45bytes), and stores concatenated values $\sigma_d || $``\_''$ || t$.

\subsection{Implementing \sys on other devices} \label{sec:other_devices}

\begin{figure}[t]
  \centering
  \includegraphics[width=\columnwidth]{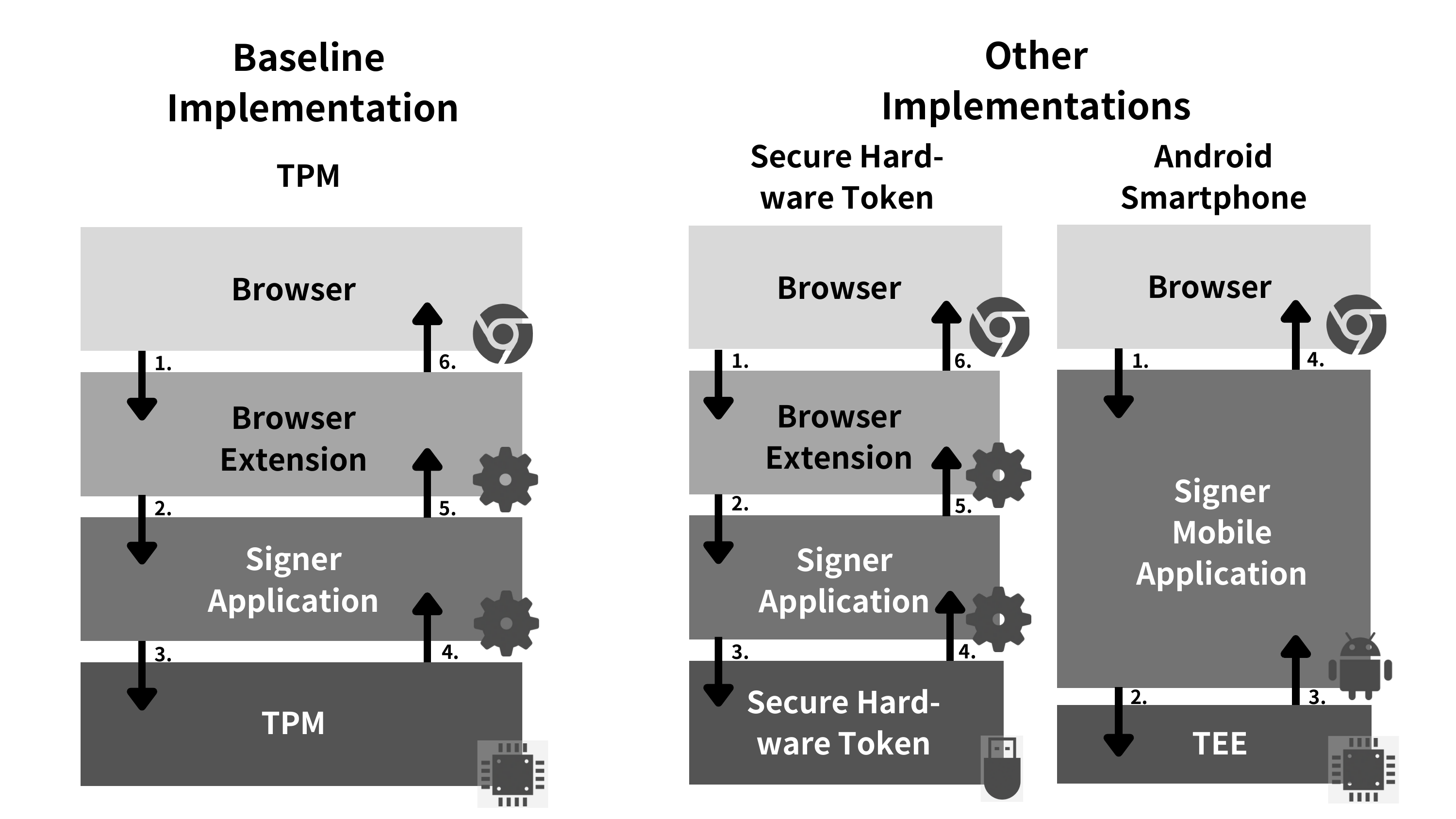}  
  \caption{Implementation of signer-side components and their relationship.}
  \label{fig:web-extention-flow}
\end{figure}

\begin{table}
  \caption{Comparison of the type of implementation}
  \label{table:compare-implemantation}
  \centering
  \begin{tabularx}{\columnwidth}{X X X X}
    \toprule
    & Baseline & Hardware Token & Smartphone \\
    \midrule
    
    \rowcolor{Gray}    
    Device & PC (carrying TPM) & Hardware Token & Android Phone \\
    
    Private key storage & TPM & Hardware Token & Encrypted File using TEE \\
    
    \rowcolor{Gray}    
    Unique Resource & Endorsement Key & Manufacturer-Installed Private Key & Serial Number, IMEI/MEID *1 \\ 
    
    Method of Proving Unique Resource & TPM Attestation & Proof of Possession of Private Key & Android ID Attestation (ARM TrustZone) \\ 
    
    \rowcolor{Gray}    
    Rate-limiting Depends on Device Security & No & No & Yes \\
    
    \bottomrule 
  \end{tabularx} \\ 

  *1 Serial Number of Phone, IMEI (International Mobile Equipment Identifier) and MEID (Mobile Equipment IDentifier)
\end{table}

TPM was our device of choice when implementing \sys.
    However, signers may only have access to computers that do not have TPM or not even have access to computers at all.
    To accommodate such users, we implemented \sys on two other devices: a secure hardware token and a smartphone.
    Although we had to make slight modifications to the specifications and the software stack during the implementation, this shows the versatility of \sys and its potential for being deployed to many users.
    Figure~\ref{fig:web-extention-flow} provides an overview of the implementation with respect to the relationship of each component regarding the different devices.

\subsubsection{Secure hardware token} \label{sec:token_impl}
We used Solokey Hacker~\cite{Solokey:online} for this implementation.
    The token includes an STM32L432KC microprocessor with an Arm Cortex-M4 MCU running at 80 MHz, 64 kB of RAM, 256 kB of flash memory, and a true random number generator.
    Solokey supports the FIDO API~\cite{SolokeyCode:online}, and we extended the Client to Authenticator Protocol (CTAP) API to support the FIDO ECDAA~\cite{FidoEcdaaBasename:online} functionalities.
    We chose to use the Xaptum ECDAA library~\cite{XaptumEcdaa:online}, as it is compatible with FIDO ECDAA version 1.1~\cite{XaptumEcdaaCompatibility:online}. 
    The library uses the Miracl AMCL cryptographic library~\cite{XaptumAmcl:online} for the necessary BN256 curve.
    Our custom Solokey firmware was written in C language.

Solokey exposes a custom \texttt{ECDAA\_SIGN} API through a modified FIDO CTAP1 (U2FHID) API that takes \texttt{message} and \texttt{basename} as inputs and outputs an ECDAA digital signature over the two pieces of data.
    To simplify the implementation process, we assume that the secure hardware token has already run the ECDAA Join protocol and has obtained a valid $cred$.

\textbf{Challenges and solutions.}
The challenge of implementing \sys on secure hardware tokens is that no tokens support FIDO ECDAA in the wild, including Solokey.
    This is a drawback, since it does not enable users to purchase a token and start using \sys immediately.
    However, we presume that some tokens will support ECDAA in the future, which we simulated by adding ECDAA functionalities to the FIDO software stack on Solokey.

Another challenge is that the FIDO ECDAA specification does not allow the use of basenames when producing digital signatures to ensure signer privacy~\cite{FidoEcdaaBasename:online} (see Section~\ref{sec:security_eval} for more discussion).
    This required the \sys implementation on the token to slightly deviate from the official specification, which, again, prevents this implementation from being used in the real world.
    Nonetheless, the Xaptum ECDAA~\cite{xaptumec40:online} implementation supports the usage of basenames, allowing us to implement the PoC.
    
\textbf{Security of device.}
Although current secure hardware tokens share the same public/private key pair to ensure device privacy~\cite{FIDO11FullBasicAttest:online}, we envision tokens used in \sys to have unique key pairs per device.
    This is possible because ECDAA allows such tokens to achieve the same level of privacy provided to current tokens.
    Moreover, with this assumption, we can guarantee the rate-limiting functionality of \sys even if per-device unique key pairs get leaked from tokens, as GMs can reject multiple Join requests from the same public key.

\subsubsection{Smartphone} \label{sec:smartphone_impl}
We used an Android~\cite{Android:online} smartphone for this implementation.
    The signer application was implemented as an Android mobile application (hereafter referred to as ``Signer mobile application'').
    The main Signer mobile application was written in Kotlin~\cite{AndroidKotlin:online}, while the ECDAA library and other cryptographic schemes were written in Golang due to security and efficiency reasons.
        Since Golang is not supported natively on Android, we used Gomobile~\cite{gomoble:online} to convert these libraries. 
    The Signer mobile application also used Android Room~\cite{Room:online} and SQLite to store its log.

\textbf{Challenges and solutions.}
The challenge of implementing \sys on a smartphone is that most mobile browsers do not support browser extensions.
    To overcome this, we used a custom URL to communicate between the browser and the Signer mobile application. 
    Briefly, the URL allows the browser to open the Signer mobile application and share the necessary information.
    Similarly, the application jumps back to the browser by following a callback URL, which allows the browser to send the generated proof back to the verifier server.
    Note that the user must check the callback URL to prevent generating proofs for unintended basenames.

Another challenge is storing private keys securely.
    This is an issue for \sys because Android does not have any API that supports the storage and usage of ECDAA keys within hardware, unlike TPM.
    We overcame this by encrypting the ECDAA private key using standard methods and decrypting it when needed.
        This way, the encryption key can be stored securely using the Android Keystore API~\cite{KeyStore:online}, which uses a TEE for storage. 
    However, it is worth noting that this requires the plaintext ECDAA private key to briefly exist on the smartphone memory.
        Although this occurs only for a short time, it makes this implementation less secure against side-channel attacks compared to the baseline implementation.
        We envision that this will no longer be an issue once the Android Keystore API supports ECDAA signing operations within a TEE.

An additional challenge is integrating smartphone identification schemes to \sys.
    This is required as the smartphone acts as a source of uniqueness for \sys, similar to attestation keys in TPM and secure hardware tokens.
    Although we could potentially use Android ID Attestation\footnote{Android ID Attestation relies on ARM TrustZone for protecting resources proving device uniqueness. Thus, adversaries may steal this resource via side-channel attacks. Table~\ref{table:compare-implemantation} reflects this fact.}~\cite{AndroidID:online}, integrating it into the ECDAA Join protocol is no easy task.
    Since this requires a large implementation effort, we leave this for future work.
    Therefore, similar to the secure hardware token implementation, we assume that the smartphone has already obtained a valid $cred$.

\textbf{Security of device.}
The Android ID attestation documentation does not clearly describe whether the public/private key pair used to generate the proof of the device serial number is unique to the device.
    However, the fact that the public key is not listed in the identifying information in the documentation leads us to believe that the key pair is shared across multiple devices.
    Therefore, once an adversary can extract the private key protected by ARM TrustZone via side-channel attacks, they can forge valid proofs of identifiers of phones that do not exist.
    We, thus, determine that the rate-limiting functionality of \sys using Android phones depends on the security of the TEE.

\section{Performance Evaluation}\label{sec:eval}
This section presents the latency, bandwidth, and storage evaluation results of the proposed system.

For the TPM and secure hardware token evaluation, we used a Lenovo ThinkPad A285~\cite{ThinkPad18:online} with an AMD Ryzen 5 PRO 2500U CPU (4 cores, 2 GHz) with 16 GB of RAM, representing a standard consumer-grade laptop.
    For this evaluation, the signer application ran within a Docker container.
    The containers ran Ubuntu 22.04 and utilized standard Docker APIs to mount necessary hardware devices to the signer container.
    For the TPM it was \texttt{/dev/tpm} and for the secure hardware token it was either \texttt{/dev/ttyUSB0} or \texttt{/dev/ttyACM0}.
For the smartphone evaluation, we used an Android Pixel 6a smartphone with a Google Tensor CPU ($2 \times 2.8$ GHz, $2 \times 2.25$ GHz, $4 \times 1.8$ GHz) and 6 GB of RAM, running Android version 13.
    The signer application ran as a smartphone application.
All verifier-side experiments ran on an AS-2024US-TRT/3Y machine with two AMD EPYC 7513 CPUs (32 cores, 2.6 GHz) with 512 GB of RAM, representing a standard server-grade machine. 
Unless otherwise stated, the numbers reported are an average of 20 measurements.

\subsection{Latency Evaluation}
\subsubsection{Baseline latency evaluation} \label{sec:baseline_eval}
We first present the proof generation latency of the three devices as well as the proof verification latency under simulated baseline conditions.
    For the setup, we do not populate the signer-side log with any entries, simulating a situation where a user has not visited websites with \sys.
    On the other hand, the verifier-side log is populated with 100,000 entries to simulate a situation where the verifier has been running for some time.
The results are presented in Table~\ref{table:e2e_latency}.

\begin{table}[t]
  \caption{End-to-end latency evaluation of \sys (Cryptographic operation + Database operation)}
  \label{table:e2e_latency}
  \centering
  \begin{tabularx}{\columnwidth}{X X X}
    \toprule
    \multicolumn{2}{c}{Proof generation}  & Proof verification                   \\ \midrule
    Hardware Device & Measurement  [ms]   & Measurement [ms]                     \\ \midrule
    TPM (baseline)  & 243.16 (236 + 7.16) & \multirow{3}{*}{84.1 (73.7 + 10.4)}  \\ 
    Hardware Token  & 2771 (2764 + 7.16)  &                                      \\ 
    Smartphone      & 26.4 (23 + 3.4)     &                                      \\ 
    \bottomrule 
  \end{tabularx}
\end{table}

For the proof generation phase, we can see that the smartphone is the fastest, taking 26.4 ms (cryptographic operation: 23 ms, database operation: 3.4 ms), the TPM the second fastest, taking 243.16 ms (236 ms, 7.16 ms), and the hardware token the slowest, taking 2771 ms (2764 ms, 7.16 ms).

It is interesting to observe that the TPM exhibits a speed that is an order of magnitude faster than that of the secure hardware token when generating signatures, despite both devices having processors with similar frequencies.
    We hypothesize that this is because TPM has optimized hardware for producing DAA signatures, whereas the secure hardware token does not.

The average latency for the proof verification phase is 84.1 ms, with the verification operation taking about 73.7 ms and the database lookup and insert taking 10.4 ms.
    This is an acceptable latency since the AMCL cryptographic library benchmark test results on the same server-grade machine were in the same order of magnitude (i.e., tens of milliseconds).
    Additionally, \cite{xi2014direct} reports similar numbers for the DAA verification process on a CPU with a faster clock frequency.
    Note that the verification operation currently uses only one CPU core and could benefit from muti-threading as well as standard load-balancing techniques when dealing with a large number of verification requests.

\subsubsection{Latency evaluation under extreme conditions} \label{sec:extreme_latency}
Next, we evaluated the latency of proof generation and verification phases under certain extreme conditions.
    For the proof generation phase, the signer-side log was populated with a large number of entries, and the verifier was required to check a revocation list with a large number of entries for the proof verification phase.
    
The results are presented in Table~\ref{table:extreme_latency}.
    Differences between Table~\ref{table:e2e_latency} and~\ref{table:extreme_latency} are highlighted in bold text.
    We observe that while the overhead of having a large signer-side log is minimal, having a large revocation list introduces a large overhead.
    A thorough analysis of these results is shown below.

\begin{table}[t]
  \caption{End-to-end baseline latency evaluation of \sys under extreme conditions (Cryptographic operation + Database operation). Differences are highlighted in bold.}
  \label{table:extreme_latency}
  \centering
  \begin{tabularx}{\columnwidth}{X X X}
    \toprule
    Condition             & Proof generation [ms]                 & Proof verification [ms]              \\ \midrule
    Large signer log      & \textbf{243.23} (236 + \textbf{7.23}) & 84.1 (73.7 + 10.4)                   \\ 
    Large revocation list & 243.16 (236 + 7.16)                   & \textbf{151.4} (\textbf{141} + 10.4) \\ 
    \bottomrule 
  \end{tabularx}
\end{table}

\taggedpara{Latency of signing with a large number of signer logs}
This evaluation simulates where a signer visits a large number of websites that support \sys.
    For this experiment, we populated $Log_s$ with 1,000 entries and measured how long it took to generate a signature.
    1,000 entries are significantly larger than the number of distinct web pages\footnote{Note that ``distinct web page'' refers to a unique URL and not necessarily a unique domain, i.e., the actual number of distinct domains may be smaller.} visited by a user per day (163 web pages)~\cite{crichton2021home}, representing an upper bound.

The average latency for this measurement was 243.23 ms.
    The signing operation took 236 ms, while database lookup and insertion took 7.23 ms.
    Comparing this to the results presented in Section~\ref{sec:baseline_eval}, we can see that the database operation latency increased by 0.07 ms, which is negligible.
    This shows that the signer-side latency does not get affected even if signers visit a large number of websites.

\taggedpara{Latency of verifying with a large number of revocations}
This evaluation simulates where many signers had their group private keys revoked.
    The number of entries in the revocation list was set to 50, which is the number of maximum allowed entries in a revocation list in EPID~\cite{ruan2014privacy}.

The average verification latency with 50 entries in the revocation list was 151.4 ms, with the verification operation taking 141 ms and the database lookup operation taking 10.4 ms.
    This is nearly two times slower than the latency reported in the baseline evaluation, although the overall end-to-end latency is still within the order of hundreds of milliseconds.
    Even though it is highly unlikely that there are this many entries in the revocation list, this shows that verifiers must take into account such a case when designing their services.

\subsubsection{Comparison of latency results}
Overall, we conclude that \sys introduces minimal latency overhead, even under extreme conditions.
    Table~\ref{table:latency_comparison} compares the latency of \sys with the numbers reported in related work.
    While evaluation environments vary across different papers, we can observe that the latency does not drastically differ (except for Opaak).
    It is worth noting that our baseline evaluation uses a resource-limited TPM, whereas others use more powerful devices.

\begin{table}[t]
  \caption{Comparison of latency measurements between \sys (highlighted in bold) and related work.}
  \label{table:latency_comparison}
  \centering
  \begin{tabularx}{\columnwidth}{X X p{1.25cm}}
    \toprule
    Work                                                   & Proof generation [ms]        & Proof verification [ms]     \\ \midrule
    \textbf{\sys (TPM, baseline)}                          & \textbf{243.16}              & \textbf{84.1}               \\ 
    \textbf{\sys (Hardware Token)}                         & \textbf{2771}                & \textbf{84.1}               \\ 
    \textbf{\sys (Smartphone)}                             & \textbf{26.4}                & \textbf{84.1}               \\ 
    CACTI~\cite{nakatsuka2021cacti}                        & 211.9                        & 27.3                        \\ 
    Privacy Pass (N tokens)~\cite{davidson2018privacypass} & 341.48 + 180.87 $\times$ N   & 57.8                        \\ 
    Opaak~\cite{maganis2012opaak}                          & \multicolumn{2}{c}{2550 (combined measurement)}            \\ 
    \bottomrule 
  \end{tabularx}
\end{table}

\subsection{Bandwidth Evaluation}
We measured the amount of data transferred between the signer and verifier when using \sys.
    Recall that the verifier sends $t$ to the signer when requesting the rate-assuring proof, and the signer responds by sending $\sigma$ to the verifier.

The size of $t$ is 19 bytes, which is a 64-bit Unix timestamp encoded as a string.
The size of the raw signature $\sigma$ is 261 bytes, consisting of five 33-byte elliptic curve points and three 32-bit big integers.
    Note that we use Base64 when encoding the raw data and gob (Golang encoding format) when bundling the signature parameters. Thus the actual size of $\sigma$ is 660 bytes.
    We can observe that this number is small, thus deploying \sys should only introduce minimal bandwidth overhead.

\subsection{Storage Evaluation}
Finally, we measured the amount of storage it takes to store the logs for both the signer and verifier sides.

The amount of storage required to store 1,000 entries in the signer-side log is 94.2 KB and 6.64 MB for the 100,000-entry verifier-side log.
    This indicates that our PoC stores only a small amount of data for a large number of logs and can be further optimized.

\section{Security evaluation} \label{sec:security_eval}
In this section, we provide a security analysis of \sys based on the threat model and requirements defined in Section~\ref{sec:proposal}.

\textbf{Signature forgery attacks.}
A malicious signer may attempt to generate a fake rate-assuring proof to cheat the system.
    The proposed system prevents this, as it is impossible for signers to receive a valid $cred$ from GMs without a genuine hardware security device.
    Even if the signer skips the above step altogether and generates a fake $cred$, it will still get caught since the $cred$ used to generate the rate-assuring proof will not be valid.

\textbf{Timestamp forgery attacks.}
A malicious signer may attempt to use a $t$ that the verifier did not provide.
    For instance, if the valid $t$ is \texttt{10:00-11:00}, the malicious signer may try to create a rate-assuring proof on \texttt{10:00-10:10} in an attempt to get more access to the verifier's resource.
    This is preventable by the verifier, since the proof provided by the signer will include a different $t$ than the one provided.

\textbf{Generating multiple proofs upfront.}
A malicious signer could generate multiple rate-assuring proofs with different $t$ upfront and send them all at once to the verifier.
    However, this will not grant the signer with the same amount of access as the number of generated proofs, since the $t$ in the proof will not match the one expected by the verifier and will be rejected.

\textbf{Obtaining multiple credentials.}
A malicious signer may attempt to generate multiple $usk$s and obtain credentials over them, allowing the signers to generate multiple rate-assuring proofs for a given set of $t$ and $bsn$.
    This is impossible, as the GM will know that the same signer is requesting multiple credentials via their attestation key (i.e., $EK$).

\textbf{Device Reset attacks.}
Malicious signers can always reset their hardware security device and re-run the Join protocol claiming that they are a fresh device.
    The proposed method prevents this by having the GM limit the number of times a device can run the Join protocol.
    This is possible because the GM can keep track of the number of times it has observed the same $EK$.

\textbf{Signer tracking via proofs.}
A malicious verifier may attempt to track signers using their rate-assuring proofs in the following two ways: 
    (1) linking two proofs from two periods in time; and/or
    (2) colluding with another verifier and linking proofs that were sent to the two verifiers.
    These attacks are not possible because using either a different $t$ or $bsn$ will generate a signature that is indistinguishable from other signatures.

\textbf{Signer tracking via $t$.}
A malicious verifier may attempt to track signers using $t$ which they provide during the proof generation phase.
    For instance, a verifier may provide a signer of interest with $t'$ while providing the rest with $t$.
    Sometime in the future, a verifier suspects that the same signer is requesting access, and provides the suspected signer with $t'$.
    If the signer cannot respond with the proof, the verifier's suspicion is correct and has successfully identified a signer.
    This is easily detectable since $t'$ is from the past.

Similarly, a malicious verifier may provide a signer with a $t$ which is valid for an extremely long time.
    If a verifier suspects that the same signer is requesting access, they provide the signer with $t'$ that overlaps with $t$.
    If the signer is in fact the same, they cannot generate proof, allowing the verifier to confirm its suspicion.
    This is also detectable since the signer observes that the verifier is providing overlapping time windows, which violates the protocol.

In either case, signers should report such malicious activities to third-party auditors so that malicious verifiers are held accountable.

\textbf{The use of basenames in \sys and its privacy implications.}
We have discussed in Section~\ref{sec:token_impl} that the FIDO ECDAA specification prohibits the use of basenames due to privacy concerns.
    This is because FIDO ECDAA aims to provide full anonymity to its users and the use of basenames may cause signatures to be linked to each other (i.e., pseudonymity, see Section~\ref{sec:background:daa} for more details).

This pseudonymity property is necessary for \sys as it is what enables verifiers to know that a certain signer is conducting multiple actions within a given time window.
    Although this may seem that \sys is unable to provide full anonymity to its signers, this is not the case.
    This is because malicious signers are the only entities that produce signatures that can be linked to each other.
    On the other hand, legitimate signers do \emph{not} send multiple rate-assuring proofs within the same time window.
    In addition, the use of $Log_s$ during the proof generation phase described in Section~\ref{sec:protocol} allows legitimate signers to keep track of whether they have generated proofs for the same verifier within the same time window.
    Therefore, as long as signers abide by the \sys protocol they can maintain full anonymity.

\textbf{Violating signer privacy via side-channel attacks.}
Malicious actors may attempt to steal the $usk$ from signers using side-channel attacks.
    This is an issue since once an attacker steals the $usk$, they are able to link rate-assuring proofs to an individual signer, by calculating the pseudonym used by the signer.
    Although it is not possible to prevent this via cryptographic approaches, the use of secure hardware in \sys significantly increases the difficulty of the attack for adversaries (except for the smartphone implementation, see Section~\ref{sec:impl} for more details).

\textbf{Network attacks.}
A malicious party on the network may change the values for $t$ and/or $bsn$ in an attempt to mount a DoS attack against an unsuspecting signer.
    This is preventable since the signer checks whether the $t$ and $bsn$ are valid before generating the proof.
    Furthermore, an honest signer will detect that there is a network adversary, as their requests will be denied despite legitimately producing the proof.

\textbf{Compromised Devices.}
Malicious signers may attempt to extract the $EK$ and/or the $usk$ from hardware security devices owned by other signers through various means, including side-channel and physical attacks.
    We investigate the consequences of an adversary obtaining the two keys.

\textit{Leaked $EK$.}
A large number of devices may have their $EK$ extracted while in possession of other parties after manufacturing (e.g., during shipping).
    This allows a malicious signer to generate valid rate-assuring proofs for a given $t$ and $bsn$, granting them an abnormal number of accesses.
However, this is \emph{detectable} by the victim, as a GM will refuse to issue a $cred$ as it has already done so to the same $EK$ in the past.
    Moreover, even if a GM issues a valid $cred$ to the victim, the leaked $EK$ does \emph{not} violate the privacy of the victim, since the $usk$ generated by a victim cannot be linked to $EK$.

\textit{Leaked $usk$.}
Extracting $usk$ from the hardware security device allows the adversary to not only generate valid rate-assuring proofs but also violate the privacy of the victim.
    This is because the adversary can link victim-generated proofs to $usk$.
However, extracting $usk$ is extremely challenging for the adversary.
    First, the adversary is required to have access to the victim's device (e.g., via malware),
        This is because the victim generates $usk$ only when the device is within their possession.
    Second, the hardware security device stores $usk$ within its tamper-proof memory.
        This would be even more challenging if the device is detachable (e.g., secure hardware token).
In essence, the tamper-proof assumption of the hardware security device can be seen as a means to protect the \emph{signer privacy} property of \sys.

\textit{Leaked $EK$ and $usk$.}
The unforgeability and unlinkability properties of \sys would be violated if both $EK$ and $usk$ are stolen.
    However, we would like to emphasize that even in this situation, the \emph{rate-limiting} property remains intact.
    The intuition behind this is that the verifier will accept a proof generated by \emph{either} the adversary \emph{or} the victim, but not both.

\textbf{Rogue GMs.}
A malicious GM may attempt to compromise signer privacy by colluding with a malicious verifier.
    However, this is prevented by the use of DAA, as it is impossible to link $EK$ and $usk$, even by the GM.

A malicious signer may also attempt to produce multiple rate-assuring proofs within the same period of time for the same verifier with the help of a malicious GM.
    This is done by having the malicious GM produce credentials for the multiple $usk$s generated within the single hardware security device.
    However, once caught (e.g., through community reporting), verifiers will stop verifying proofs that use $gpk$s issued from these GMs, essentially losing the trust of the public.
    Therefore, we assume that a \emph{rational} GM will not engage in such activities and will be heavily penalized if discovered.

\textbf{GM with only one member.}
It is easy to imagine that malicious verifiers will be able to track a signer that generates proofs that are verifiable using a $gpk$ of a group of which they are the sole member.
    Since \sys does not regulate such behavior, it cannot prevent such incidents from happening.
    Signers must be aware of the GM they are interacting with and should consider joining a group that is reputable and popular.
    GMs can use this information to stand out from other GMs, for example by publishing the number of current group members.

\taggedpara{Summary}
Overall, we claim that the proposed system successfully meets the security requirements defined in Section~\ref{sec:proposal}.
    Specifically, the \textbf{Unforgeability} requirement is satisfied since it is impossible for the signer to generate rate-assuring proofs without a genuine device and using valid $t$ and $bsn$.
    The \textbf{Unlinkability} requirement is also satisfied because of the use of DAA and the rate-assuring proofs not leaking any information regarding the signer's identity.
    The \textbf{Rate-limiting} requirement is met because rate-assuring proofs that carry invalid timestamps will not be accepted by the verifier, thus preventing any access above the rate set by the verifier.
    Moreover, any signer attempting to collude with a malicious GM will be detected and punished.

\section{Usability Analysis} \label{sec:usability_analysis}
In this section, we provide an analysis of \sys from the perspective of signer usability.

\textbf{Installation.}
Installing \sys on the signer platform requires the \sys browser extension and native application to be downloaded and installed.
    We implemented \sys to ensure that the installation process does not impose excessive overhead on the user.
        However, the process can be further improved by integrating the extension logic and native application functionalities directly into the browser.
        This way, \sys{'s} signer-side logic will work ``out of the box'', providing a seamless user experience.

\textbf{Signer-perceived latency.}
Minimizing the time a signer needs to spend when using the system is crucial for its usability.
    As shown in Tables~\ref{table:e2e_latency} and \ref{table:latency_comparison}, the end-to-end latency of \sys ranges between 100 and 2800 ms, comparable to other state-of-the-art rate-limiting systems.
    Therefore we claim that the latency introduced by \sys is minimal and does not degrade user experience.

\textbf{Signer involvement.}
\sys is designed to be fully transparent to the signer, i.e., signer input is not required when creating rate-assuring proofs.
    However, there are cases where signer input may be useful.
    
For instance, a malicious verifier may use $t$ to track signers (see Section~\ref{sec:security_eval}).
    Such an attack can be automatically detected via the aid of signer-side software (e.g., \sys browser extension).
    However, there may be cases where the software is not confident enough to either reject or accept a $t$.
    In such a case, signer input would be required to make the final decision.

Another example is malware on signer-side devices.
    This allows malicious signers to use other legitimate signers' devices to generate valid rate-assuring proofs (i.e., cuckoo attack).
    Preventing such an attack would require signer confirmation via I/O which is hardwired to the hardware security device (e.g., secure I/O in ARM TrustZone).

In short, incorporating signer input improves security.
    However, requiring frequent input degrades user experience due to UI fatigue.
    To strike a balance between user experience and security, we envision signer-side software allowing signers to choose from the following options:
        (1) \emph{Ask every time}: Requires signer input each time it generates a proof (default).
        (2) \emph{Ask for untrusted}: Requires signer input only for websites that are not trusted by the signer.
        (3) \emph{Do not ask}: Does not require any signer input.

\textbf{Usability of supported devices.}
Supporting many devices is crucial for adoption, as it allows signers to use their preferred device.
    Here, we assess the usability of the three devices used in this paper.
    Signers that want absolutely no involvement during the proof generation phase are recommended to use devices with TPM.
    Signers comfortable with minimal user involvement may opt to use smartphones.
    Secure hardware tokens require high signer involvement as they need to be plugged in every time.

\taggedpara{Summary}
Overall, we claim that \sys is designed and implemented with usability in mind.
    The installation process is minimal, with potential optimization, and signer-perceived latency is comparable to other related work.
In addition, signer involvement in \sys is optional.
    While important when mitigating some attacks, we have demonstrated that the level of involvement can be adjusted according to the signer's preference.
Furthermore, signers have a wide selection of devices to choose from when participating in the \sys system.

\section{Discussion} \label{sec:discussion}

\subsection{Fallback options}
Signers who cannot participate in the \sys protocol due to lack of access to applicable devices or do not want to participate in the protocol will fall back to the current rate-limiting services provided by the verifier.
    Note that this does not change the current state of security and privacy provided to the verifiers and signers.
    Moreover, as we have shown in Section~\ref{sec:impl}, \sys can be implemented and deployed on various types of devices.
    We are certain that users will have at least one device compatible with \sys.

\subsection{Deployment considerations and incentives}
\subsubsection{CDN and 3rd party integration}
Although \sys was designed in mind to reduce deployment effort by using well-known components (e.g., DAA, SQLite, Golang), we expect that verifier operators may not be familiar with the underlying technology stack used in \sys.
    To accommodate such needs, we envision Content Delivery Networks (CDNs) and other 3rd parties operating the \sys verifier on behalf of its customers.

CDNs host the content of their customers in servers distributed across the globe, reducing delivery latency to users.
    Recently, CDNs started to offer security services to their customers, especially protection against abusive actions.
    The three most popular CDNs -- Cloudflare~\cite{CloudflareRateLimiting:online}, Akamai~\cite{AkamaiRateLimiting:online}, and Fastly~\cite{FastlyRateLimiting:online} -- all offer rate-limiting services, using various methods ranging from basic threshold-based limiting to CAPTCHAs.
    \sys can be easily integrated into the CDN gateways, providing stronger rate-limiting capabilities for customers who are interested.

\subsubsection{Website operator incentives}
Websites have several incentives to integrate \sys into their services.
    First of all, \sys provides strong guarantees of rate-limiting through proven cryptographic schemes.
    Secondly, \sys requires minimal latency when generating and verifying rate-assuring proofs, limiting the impact on user experience.
    Thirdly, the unlinkability property of \sys enables users to participate in a private manner.

Website operators can leverage \sys{'s} privacy aspect and promote their commitment to safeguarding users' privacy.
    This can have a positive impact on their service, as it may attract additional privacy-conscious users.

\subsubsection{GM operator incentives}
The role of the GM in \sys is to check the genuineness of the hardware security device and issue a credential if confirmed.
    Since this only occurs once every lifetime of a hardware security device (unless it gets reset, see Section~\ref{sec:security_eval} for more information about this), the operation of the GM is relatively lightweight.
    Various organizations could run a GM, each with different incentives, such as online identity providers providing federated login services and non-profit organizations (e.g., Let's Encrypt~\cite{LetsEncrypt:online}).

\subsection{Supporting multiple GMs} \label{sec:multi_gm}
So far, we have not considered more than one GM in the \sys ecosystem, but given the incentives shown above, many GMs may want to benefit from \sys.
    \sys does not put a limit on the number of GMs it supports.
    However, this allows a malicious signer to obtain multiple credentials for a pair of $upk$ and $EKCert$.
    This is an issue since malicious signers would be able to generate multiple rate-assuring proofs that are unlinkable to each other.

Depending on the services they operate, verifiers may want to choose one of the following solutions:
    (1) For verifiers that cannot allow more than one access to a certain resource per signer at a given point in time (e.g., online ticket sellers), they may want to notify the signers that they will only accept signatures from certain single GM to prevent such issue; or
    (2) For verifiers that have some tolerance, they may want to opt to support as many GMs as possible to attract signers.

It is easy to imagine that verifying the proof one by one against a list of GM public keys is not efficient.
    Therefore, once support for multiple GMs in the ecosystem is introduced, we propose to make the following modifications to the proof generation phase:
    (1) When submitting the rate-assuring proof to the verifier, the signer must also include the identifier of the GM (e.g., the domain name).
    (2) The verifier must maintain a one-to-one mapping of GM identifiers and their group public keys, which allows efficient lookup and verification of the proof.
    (3) When the verification fails, verifiers can either choose to cross-check the proof against all public keys within the table or deny access to the service.

\subsection{Different TPM implementations}
There are four types of TPMs: discrete, integrated~\cite{IntegratedTPM:online}, firmware-based~\cite{raj2016ftpm}, and virtual/software-based~\cite{perez2006vtpm}.
    Discrete TPMs are dedicated cryptoprocessors separate from the CPU and are normally plugged into the motherboard.
    Integrated TPMs are CPUs that provide functionalities and security guarantees similar to those of TPMs without a dedicated processor.
    Firmware TPMs (fTPMs) use TEEs to store necessary secret keys and to protect the integrity of important data.
    Virtual/Software-based TPMs are a full software implementation of TPM functionalities and do not utilize hardware to store secrets nor carry valid proof of uniqueness.

Discrete and integrated TPMs satisfy the four requirements discussed in Section~\ref{sec:proposal}, while firmware and virtual/software-based TPMs do not satisfy some requirements.
    In particular, fTPMs cannot protect secret keys such as $EK$ from side-channel attacks due to using TEEs (e.g., \cite{jacob2023faultpm}).
        However, fTPMs can still be used in \sys, as the underlying TEE acts as a source of uniqueness.
On the other hand, virtual/software-based TPMs do not provide proof of uniqueness and, thus, do not have valid $EK$. 
    Therefore, they would not be able to be used in \sys.

\subsection{Balancing deployability and rate-limiting functionality} \label{sec:deployability_and_rate-limiting}
Supporting different types of hardware security devices is crucial for deployability so that services would be more willing to employ \sys.
    As we have shown in Section~\ref{sec:impl}, \sys can be deployed on a wide range of devices.
However, there is a clear trade-off between increasing deployability and providing rate-limiting functionality.
    This is because as more devices support \sys, the easier it is for an attacker to operate farms that employ a large number of such devices.
    In this section, we take a closer look into device farms and their effect on \sys.

Similar to CAPTCHA farms, a device farm is an operation that uses a large number of devices to generate numerous valid rate-assuring proofs, attempting to attack the rate-limiting capability provided by \sys.
    Although \sys does not (and cannot) prevent such farms from being operated, it certainly can discourage them from forming.
    This is because (1) each device can only produce one proof for the same $t$ and $bsn$, and (2) the cost of purchasing such a large amount of devices would be significantly higher than other alternatives.
    For example, some CAPTCHA farms charge as little as $0.5$ to $1$ USD for solving 1,000 text-based CAPTCHAs and $3$ USD for 1,000 reCAPTCHAs~\cite{Top10Captcha:online}.
    In comparison, TPM chips usually cost around $20$ USD each, requiring the adversary to spend $20,000$ USD as an initial investment to provide the same number of rate-proofs. %

It is important to note that these farms are not a unique issue to \sys, as other related work also cannot defend against such farms (e.g., CACTI farms~\cite{nakatsuka2021cacti}).
    What is unique to \sys is, however, the fact that supporting a wide variety of devices for deployability allows adversaries to operate different kinds of farms.
        For instance, secure hardware tokens may be easier to farm compared to TPM chips since they use standard USB interfaces rather than manufacturer-specific ones.
        Moreover, USB hubs are extremely easy to obtain and make such tokens easier to connect to a host computer on a large scale.

To deal with device farms, we envision verifiers to accept rate-proofs according to the GM public key provided by the signer, similar to the discussion in Section~\ref{sec:virt_emu_devices}.
    For example, for actions requiring stricter rate-limiting policies (e.g., account creation, purchasing high-value items), verifiers may choose to accept proofs from devices that are more expensive and difficult to farm (e.g., firmware TPMs that cost several hundred USD per chip).

\subsection{Supporting additional devices} \label{sec:virt_emu_devices} \label{sec:non_supported_device}

Not all users may have access to the three devices we have shown in this paper.
    In this section, we discuss additional devices\footnote{Devices that cannot provide proof of uniqueness \emph{cannot} be used in \sys, including virtual TPM~\cite{perez2006vtpm,VMWareVirtualTPM:online} and virtual secure hardware tokens~\cite{VirtualFIDO:online}. However, such devices could be used if they can be cryptographically bound to a source of uniqueness shown in this section.} that may potentially be used in \sys. 
    Table~\ref{table:compare-unique-resource} compares the list of devices.

\textit{SIM cards.}
SMS authentication provides proof of possession of a certain phone number.
    Similar to Opaak~\cite{maganis2012opaak}, SIM cards can be used as a source of uniqueness in \sys.

\textit{Intel SGX CPUs.}
Intel CPUs from 6th generation to 10th generation support Intel SGX~\cite{costan2016intel}, which is a form of a TEE.
    Such CPUs have secret keys that are used to prove the genuineness of the platform as well as the code running inside the TEE.
    Using Intel SGX in \sys is fairly straightforward: prove to the GM during the Join protocol that it has not registered in the past and produce rate-assuring proofs within the TEE.

\textit{PUFs.}
Physical Unclonable Function (PUF)~\cite{PUF:nature,herder2014puf} is a one-way function that is unique per device.
    PUFs utilize the subtle differences in the physical characteristics of the components of the device to provide such uniqueness.
    Attestation of a PUF uses a standard challenge-response protocol that is similar to HMAC.
    Using PUFs in \sys is challenging because PUF keys are symmetric.
    Therefore, we envision that either the vendor of the PUF device will operate a GM or provide third-party GMs with proof of verification of responses.

Assuming that we have deployed \sys that accepts all the different devices we have listed above, several points must be taken into consideration.
    First, \sys does not (and cannot) prevent GMs from accepting a source of uniqueness.
        As some resources are easier to obtain than others, verifiers are advised to decide whether to accept rate-assuring proofs based on the signer's group affiliation.
            For example, verifiers that can tolerate high-rate access (e.g., API access, website viewing) may choose to allow signers with easily obtainable devices.
            Others may choose GMs that only provide $cred$ to hard-to-obtain devices (e.g., integrated TPMs running on Intel Xeon CPUs).
    Second, the security and privacy guarantees of devices differ.
        Some devices require the device to be trusted to provide necessary security and privacy guarantees (see Section~\ref{sec:security_eval} for more discussion). 
        Moreover, vulnerabilities of the underlying hardware may also cause further disruption to the guarantees (e.g., TEE side-channel attacks, attacks on PUF).

However, it is worth emphasizing that the security of the devices \emph{do not} impact the rate-limiting functionality of \sys.
    The reason behind this is similar to that discussed in Section~\ref{sec:security_eval}: Even if the source of uniqueness gets stolen from the device, the GM will only provide a valid $cred$ to only one device, be that the legitimate signer's or the attacker's.

\begin{table}
  \caption{Comparison of additional sources of uniqueness and their security}
  \label{table:compare-unique-resource}

  \centering
  \begin{tabularx}{\columnwidth}{X X X X X X X X}
    \toprule
     & SIM Card & Intel SGX CPU~\cite{costan2016intel} & PUF~\cite{PUF:nature}~\cite{herder2014puf} \\
    \midrule

    \rowcolor{Gray}
    Unique Resource & Phone Number & Provisioning Secret & PUF Response\\

    Method of Proving Unique Resources & 
SMS Authentication & Remote attestation & Challenge and Response Authentication\\ 

    \rowcolor{Gray}
    Rate-limiting Depends on Device Security & No & No & No \\
    \bottomrule 
  \end{tabularx}

\end{table}

\subsection{Device interoperability} \label{sec:device_interop}
Supporting many types of hardware security devices may also raise interoperability concerns.
    This is not an issue for \sys as long as the device and the signer owning the device adhere to the following restrictions:
        (1) The device satisfies the four properties shown in Section~\ref{sec:proposal}, and
        (2) The signer does not deviate from the protocol in Section~\ref{sec:protocol}.

\section{Limitations} \label{sec:limitations}

\textbf{Features other than rate limiting:}
\sys is a system that provides rate limiting.
    This implies that it cannot fully replace existing systems such as CAPTCHAs and SMS authentication.
    For instance, websites that rely on CAPTCHAs to provide human-bot distinction or SMS authentication to provide second-factor authentication cannot utilize \sys.
    However, we believe certain use cases exist where websites can benefit from \sys; for example, servers that can tolerate a certain amount of bot activity.

\textbf{Requires supported devices:}
\sys requires signers to own at least one of the devices shown in the paper.
    In theory, \sys can incorporate any device that satisfies the requirements listed in Section~\ref{sec:proposal}, and there are a wide variety of devices to choose from, as we have shown in Table~\ref{table:compare-unique-resource}.
    However, in reality, the GM is the entity that decides whether to support a particular device.
        It is highly unlikely that a GM will support every single type of device since it will increase the maintenance cost.
        Therefore, we presume that GMs will support only a handful of devices, restricting the choice of devices available for use in \sys.

\textbf{Actions are time-window-based:}
Signers who have taken action within a time window are required to wait until the next time window.
    This is by design, as it is a clear violation of rate-limiting if a signer can produce multiple valid rate-assuring proofs within the same $t$ for the same origin.
    However, waiting for the next time window introduces user friction and constitutes another limitation of \sys.

\section{Conclusion \& Future work}\label{sec:conclusion}

This paper proposes \sys, a novel rate-limiting protocol that uses unforgeable yet privacy-preserving rate-assuring proofs allowing verifiers to have strong assurance that signers are not acting abusively.
In contrast to previous approaches, the rate-limiting capability of our method does not rely on the security of the underlying hardware security device. %
Through a comprehensive security evaluation of \sys, we show that the proposed system can defend against adversaries targeted in our threat model.
We also provide an extensive performance evaluation of our proof-of-concept implementation, showing its practicality under both realistic and extreme conditions.
We identify the following as potential directions for future work:
\begin{itemize}
    \setlength{\itemsep}{0pt}
    \item Extend \sys to support EPID and other hardware security devices,
    \item Explore methods to protect secret keys from side-channel attacks,
    \item Optimize the system to further reduce latency and storage overhead, and 
    \item Investigate the deployment of \sys in real-world scenarios.
\end{itemize}

\section*{Acknowledgment}

We thank the anonymous reviewers and the shepherd for providing us with valuable feedback on prior versions of this paper.
The first author was supported in part by the Cybozu Lab Youth program.
Authors from Keio University received support from the Keio SFC Internet Research Laboratory.
The second author was supported in part by The Nakajima Foundation.



\normalsize
\raggedbottom
\clearpage

\end{document}